# Prototyping with SDR: a quick way to play with next-gen communications systems

Jorge Baranda*, Pol Henarejos*, Yan Grunenberger* and Montse Nájar*†
*Centre Tecnologic de Telecomunicacions de Catalunya (CTTC), Castelldefels, Barcelona (Spain)
†Dep. of Signal Theory and Communications, Universitat Politècnica de Catalunya (UPC)

Email: {jorge.baranda, pol.henarejos, yan.grunenberger}@cttc.es
, montse.najar@.upc.edu

*Abstract*—In this paper we present our approach regarding the implementation of new wireless radio receiver exploiting filterbank techniques, using a software-development driven approach. Since most of the common radio communications systems share a similar structure, this can be exploited creating a framework which provides a generic layout and tools to construct a reconfigurable transmitter and/or receiver. By combining the use of the Universal Software Radio Peripheral version 2 (USRP2) with a generic object-oriented framework of our own built on top of the GNU Radio software framework, we have been able to quickly implement a working proof of concept of an Uplink (UL) Filterbank Multicarrier (FBMC) receiver, both for Single-Input Single-Output (SISO) and Multiple-Input Multiple-Output (MIMO) scenario, within the project of the 7th European framework called PHYDYAS. We described here the methodology we have applied from software engineering in order to build this demonstrator, which shows the suitability of using Software Defined Radio (SDR) technologies for fast prototyping of new wireless communication systems.

## I. INTRODUCTION

In the past, real-world prototyping of communication systems were not very common due to high financial costs and long development times derived from the use of dedicated hardware (HW) such as field programmable gate arrays (FPGA), application specific integrated circuits (ASICs) and digital signal processors (DSPs). Nonetheless, this implementation is vital because they allow the exposure of problems, such as unexpected system behavior, real-world impairments and design flaws. In spite of this, HW based approach also suffers from lack of flexibility and modularity, which on the other hand are highly desirable qualities.

These qualities are provided by SDR platforms, where physical layer (PHY) and medium access control (MAC) functions are performed in software in a General Purpose Processor (GPP), while only the radio frequency (RF) and signal conversion functions such as sampling and downconversion are performed in programmable hardware. Moreover, these platforms use, as development tools, popular high level programming languages such as C/C++, which present the following advantage: as these programming languages are dominant, there is a great number of experienced programmers and a number of well-written peer-reviewed libraries are available, which open the door to create communities of software radio developers.

Until now, SDR implementations aimed to replicate already HW-implemented signal processing schemes into the software domain. But in recent years, the evolution of SDR technologies is making them a real alternative at the time to build flexible testbeds implementing communication systems with high demands on bandwidth to accommodate high data rate transmission, both for current and next generation of wireless systems such as Mobile WiMAX, 3GPP Long Term Evolution (LTE), IEEE 802.11n and the upcoming White Space Devices, were novel algorithms can be evaluated, accelerating the transition from simulation to demonstration with real radio signals.

This evolution has been possible thanks to increased capacity platform interfaces, increasingly diverse range of processors, increased on-board processing capability and improvement in the quality of RF components such as mixers and data converters. In [1], a list of different SDR platforms can be found. From all of them, the authors want to highlight the family of platforms developed by Ettus Research [2], specially the USRP2 and its predecessors, because of its trade-off between price and performance. Furthermore, these are the hardware platforms for the GNU Radio project [3]. GNU Radio project is an enormous body of pre-written, free software in continuous development by a community of programmers, who have developed blocks of code in C++ to handle a wide range of signal processing functions, as well as the blocks which interface between the USRP devices and the GPP.

*A. Related work*

Not so many papers describing SDR implementation are written from the architectural point of view, most of them either describe the technical means (FPGA, DSP) or explain in detail a specific implementation, but only a few describe the work flow process and its viability. As introduced by Mitola in the preface of [5], "Knowing how to code a radio algorithm in C on a DSP just does not give a software engineer the core skills needed to contribute effectively to software radio architecture". While this is entirely true, we can also state that the same could be applied to radio engineers when it comes to software development.

In [6], lots of the actual tools used rely on proprietary elements, which definitely reduce the possibility to simple reuse of part of the effort already done. This is partly due to use of hardware-specific accelerator, and partly due to the strict framework imposed by the avaibility of Simulink-based generator tools. In [7] there is an interesting feedback on how software engineering practices have eased the implementation process, but specifically adapt the platform to the Global Positioning System (GPS) designer point of view.

*B. Paper contribution*

Regarding all the positive aspects and the future possibilities of SDR technologies, this article aims to introduce a flexible development framework for rapid prototyping and algorithm verification of physical layer concepts named uPHYLA (Universal Physical Layer) based on the family of USRP devices and the GNU Radio software project. Using the philosophy of GNU Radio project, uPHYLA framework has been conceived and developed to make shorter *the learning curve* of GNU Radio, so researchers can use it with a basic understanding on the GNU Radio project. Mainly, this is achieved with the only use of a high level programming language (C++) instead of the intensive use GNU Radio makes of the Python scripting language. The fact that all signal processing blocks are designed as entities of the same type makes easier its integration in different radio communications systems.

Furthermore, the designed architecture of uPHYLA aims for the following aspects: efficiency, modularity, legacy, reusability and flexibility. In order to achieve these desired properties, best-practices techniques of software engineering have been taken as a reference. These techniques are going to be described in following sections.

uPHYLA framework has been proved as an effective tool when developing the proof of concept of the receiver of the UL FBMC system (both for SISO and MIMO 2x2 scenario) which includes the novelties presented within the project of the 7th European framework called PHYDYAS [4].

## II. uPHYLA FRAMEWORK DESCRIPTION

uPHYLA is intended to be a tool to implement into software the PHY Layer of any communication system to demonstrate, using a short development time, the performance of the different algorithms included in the software chain with real digitized signal. Hence, the design is oriented to create a layout where different communications subsystem can lay there. In this sense, uPHYLA is conceived as an upper architecture wrapping different libraries and modules to reproduce a radio communication system.

The herein proposed architecture contains a core that can construct each system with the set of subsystems involved in the communication process, which are fed with the samples coming from any device of the USRP family. Each subsystem is constituted as the connection of self-programmed basic blocks or the ones present in GNU Radio. uPHYLA uses the GNU Radio scheduler to distribute the samples among the involved blocks, so the software developer has not to be worried about the sampling flow of the whole process, remaining only as a task the design of the subsystems where samples are processed. Unlike GNU Radio, in the uPHYLA framework all is done in C++, avoiding the use of Python. Not only the algorithm of the processing block is coded in C++ but also its definition in the system layout and the interconnections with other blocks. In this way, uPHYLA enhances the performance of GNU Radio in terms of execution time because compiled languages (C++) run faster than interpreted languages (Python).

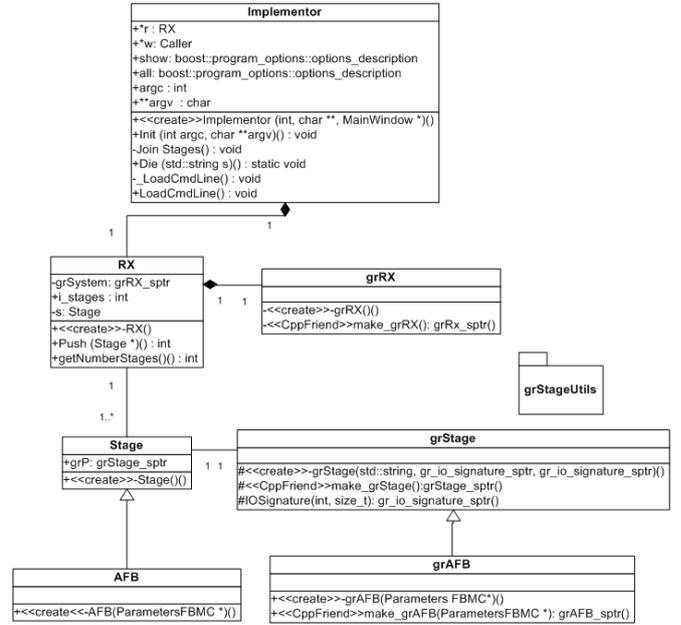

Fig. 1. Unified Modeling Language (UML) diagram of uPHYLA wrapper classes

The core of the uPHYLA framework platform resides in the wrapper files. With these files, the platform is organized into two-level architecture to make more flexible the creation and insertion of new modules and to make transparent the use of the different processing blocks of the GNU Radio framework. Wrapper files have been developed dividing the system into two types of classes. The upper level interacts with its counterpart "gr_class", making possible the use of the GNU Radio resources, as it is illustrated in the Fig. 1. In the bottom level, the different basic processing blocks (self-programmed or the ones included in the GNU Radio framework) are connected to constitute a stage in the processing chain, for instance the synchronization stage of a receiver. There is one main class, called *'Implementor'* which acts as the container and the link of the two parts of the system and implements the layout of the communication system.

Other important libraries used in the uPHYLA framework are Boost([8]) and QT/qwt([9],[10]). In GNU Radio, Boost libraries are used mostly for an efficient memory management. In the uPHYLA framework is used as well to provide the communication system of an interface to be configured by file or by command line, so it is not necessary to recompile when an option has been modified, for instance the decimation rate applied in the USRP device. Graphical interface capabilities are provided by QT/qwt libraries. These capabilities are included in the architecture in such way they can be particularized to show relevant information of each communication system.

## III. PROGRAMMING METHODOLOGY

After a brief description of the developed framework, this Section aims to explain the different software engineering techniques used to provide the framework of the capacities before mentioned. The systematic application of these kind of techniques speeds up the development process, reduce the time to solve some problems with effective proven solutions and ensures traceability, which reduces maintenance costs.

### A. Design Patterns

The idea of design patterns was introduced by Christopher Alexander in 1977 as an architectural concept. The application into the software world was pushed by [11]. Software design patterns provide solutions to a common occurring problem in software design, facilitating the reuse of successful proven designs. As a side characteristic, the use of them helps to create a common terminology which simplifies communication between the members of the development team.

As the nature of the problem they attempt to solve is different, they are subdivided into categories: creational (patterns dealing with object creation or class instantiation), structural (patterns which define relationships between elements to obtain new functionalities) or behavioral (patterns which are concerned with the communication between objects).

The effectiveness of design patterns make them indispensable at the time to develop serious software projects. In fact, GNU Radio framework incorporates many of them. This is the reason why it seems natural to adopt them in the conception of our framework. Regarding this, the implementation of the presented framework applies structural (Decorator and Adapter) and behavioral patterns (Strategy and Template Method).

### B. Software development methodology

When attempting to create efficient and high quality software, the authors think that it is very important to follow the methodology imposed by the Software Development Life Cycle (SDLC) [12]. No one can discuss that developing software is a creative process, but this has to be systematized in order to not waste this creativity energy. The **analysis** of the problem, the identification and a detailed description of the **requirements**, the definition of the **architecture** and the creation of an exhaustive **testing** plan are totally necessary before the **implementation**. Special attention have to be paid to the execution of the testing plan because it validates the fulfillment of the defined requirements.

Following a test-driven approach [13], the authors defined different levels of testing which serve to validate not only the performance of each of the communication systems which lays in the uPHYLA framework but also the robustness of the uPHYLA architecture. These different levels are:

- **Unit testing:** the aim of this test is to check if the minimum basic processing block performs its assigned functionality correctly.
- **Integration testing:** the different basic processing blocks which constitute an entity are tested together to verify if they meet the feature they are supposed to carry out. This process is repeated with an incremental aggregation of different entities.

The previous levels of testing imply only the software subsystem and have been implemented with the CppUnit framework [14], which is used also in the GNU Radio Project. The use of this tool allows running the implemented tests in an automated way and the addition of new ones very easily.

- **System testing:** at this level, the complete chain is tested, both the software and the hardware subsystems (a device of the USRP family). With these tests, it is evaluated the degree of compliance of the implementation with the specified requirements.

## IV. uPHYLA BASED DEVELOPMENT: UL FBMC RECEIVER

The uPHYLA framework accomodates the proof of concept of the UL FBMC receiver (both for SISO and MIMO 2x2 scenario) proposed in the project of the 7th European framework called PHYDYAS [4]. The aim of this project is to propose a physical layer for future radio systems based on FBMC. Although this technique presents higher computational complexity, it has better performance in terms of spectrum efficiency than present OFDM based solutions and it is better suited to the new concepts of Dynamic Access Spectrum Management (DASM) and Cognitive Radio (CR). For this purpose, a WiMAX based simulator programmed in MATLAB was developed within the project. As being WiMAX based, the developed work within this project aims to be compared with state of the art systems in order to gain acceptance within the community. The proof of concept implemented under uPHYLA framework was based upon this simulator and the main characteristics are gathered in Table IV.

TABLE I
SYSTEM PARAMETERS

| Parameter | Value |
| --- | --- |
| Carrier Frequency | 2.595 GHz |
| Number of subcarriers | 1024 |
| Bandwidth | 10 MHz |
| Sampling Rate | 10 MHz |

### A. Hardware Configuration

The host computer where all the baseband processing is performed is equipped with an Intel Core 2 Quad CPU Q9400 running at 3.2 GHz (overclocked) in combination with 4 GByte of available DDR2 RAM, whose clock frequency is 800 MHz. The host is equipped with a dual-port Gigabit Ethernet (GbE) card (Intel 82576EB controller) connected to the PCI-E x8 slot of the motherboard. This card allows the connection of each of the USRP2s to one port of the card to form the MIMO receiver, without needing the MIMO cable of the manufacturer [2], which allows to connect 2 USRP2s with a single port network card. The use of this dual-port card is justified because at the time the proof of concept was developed, the software controlling the MIMO cable was not presenting a stable release.

Each USRP2 is equipped with a RFX2400 daughter board which acts as a RF front-end for the range of frequencies comprised between 2.3 to 2.9 GHz. In order to form the MIMO receiver, USRP2s devices need to be synchronized. This synchronization is performed by means of an external 10 MHz reference clock provided by an arbitrary waveform generator. This device is used also to generate a one pulse per second (1 PPS) signal, which acts as the trigger to synchronize the received stream of samples from the USRP2 devices. These references are connected directly at the front panel of the USRP2 devices through SMA connectors.

As the proof of concept herein described only comprises the receiver, extra hardware was needed to simulate the transmitter. In the case of the SISO scenario, baseband signal was generated with the simulator developed within the PHYDYAS project. This signal was loaded into an Agilent E4438C Vector Signal Generator (ESG) to be modulated at 2.595 GHz. Then samples passed through the channel emulator (Elektrobit C8 [15]) connected to the USRP2 device. In the case of the MIMO 2x2 scenario, this basic setup presents some particularities. The simulator generates a stream of baseband samples for each antenna, so these samples are loaded into different ESGs, where one acts as a master and the other as slave and are configured and adjusted to produce an aligned signal at the output of each device. Each output of the ESG is connected to different channels of the channel emulator and then connected to USRP2 devices.

### B. Software Configuration

The UL FBMC receiver demonstrator has been developed under the 64bit version of Ubuntu 10.04 and the GNU Radio framework version 3.2.2 [3]. GNU Radio mainly provides to the uPHYLA framework of the scheduler which manages the data flow among signal processing blocks. The cited release was preferred in front of version 3.3.0 to avoid stability problems, because this version was just released when the demonstrator framework was being developed. The options interface created for this receiver allows the selection of the working mode (SISO or MIMO), the configuration of the filters and the specification of the configuration of the frame (number of symbols, number of subcarriers, number of data-slots). The graphical interface is designed with QT version 4 and qwt version 5.2.1. It displays the performance of the receiver: magnitude of the channel estimation, demodulated constellation, cumulative Bit Error Rate (BER), Signal to Noise Ratio (SNR) and Carrier Frequency Offset value (CFO) (see Fig. 4). USRP2 devices are configured by means of UHD (Universal Hardware Driver), which aims to be the unique driver for the USRP family of devices. The tests during the development were carried out with the versions released between 08/17/2010 and 11/24/2010.

### C. Demonstration

The UL FBMC receiver developed under the uPHYLA framework integrates the algorithms developed in the different workpackages of the PHYDYAS project included in [17],[18],[19],[20]. This receiver can be configured to work in SISO mode or in MIMO 2x2 mode using spatial multiplexing with Zero-Forcing (ZF) equalization. The implemented receiving chain is presented in Fig. 2.

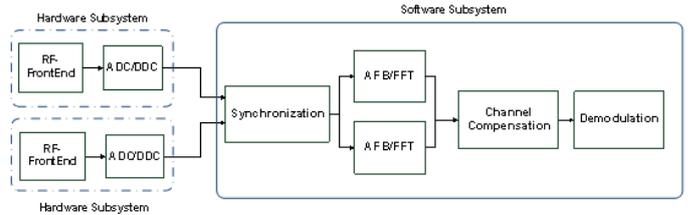

Fig. 2. Structure of the UL FBMC MIMO 2x2 receiver

The frame structure used in the proof of concept of the PHYDYAS project is depicted in Fig. 3, including a detail of the data and the pilot carriers allocation. The uplink zone covers 15 OFDM symbols in the SISO scenario, and 12 OFDM symbols in the MIMO scenario. In both scenarios, the pilots are scattered through 12 slots of allocated data using Adaptive Modulation Coding (AMC) as the pilot permutation scheme.

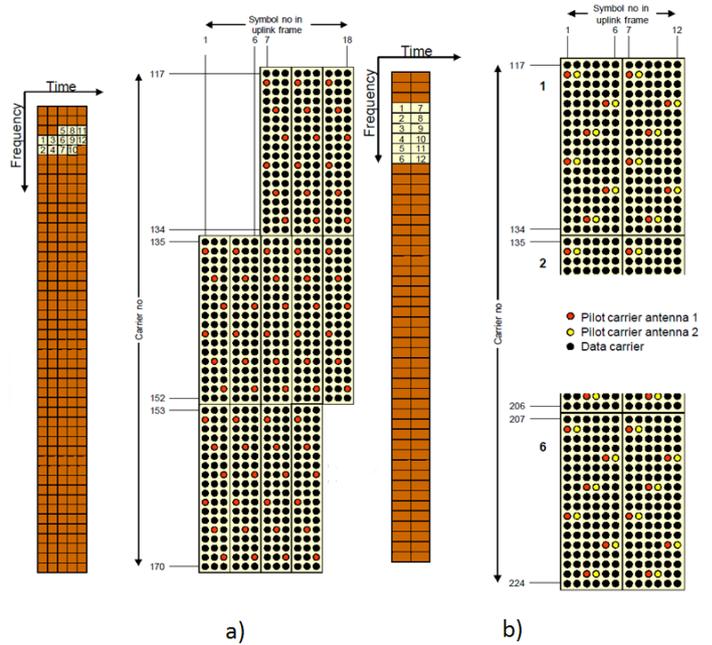

Fig. 3. UL FBMC receiver frame Structure. a) represents the SISO scenario and b) represents the MIMO 2x2 scenario.

It is important to remark that the frame is completed with a preamble, which is not shown here. This preamble is needed because the implementation only covers the uplink receiver.

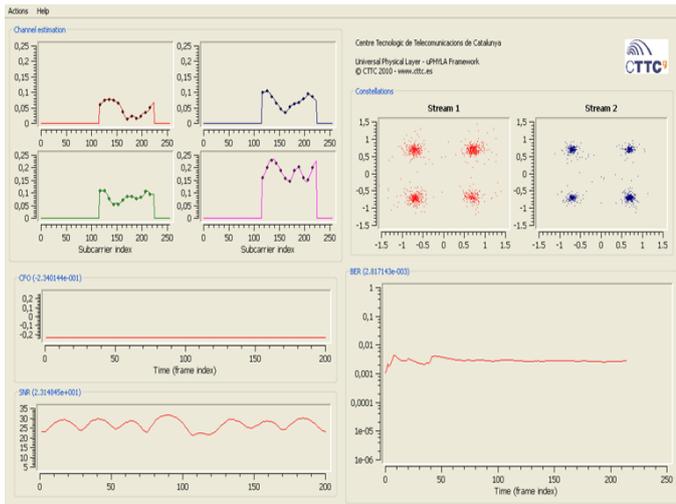

Fig. 4. Performance of the UL FBMC MIMO 2x2 receiver

Normally, this preamble should not be needed because the synchronization to the base station is performed during the connection setup by means of a two-way process.

After the signal is synchronized, the Analysis FilterBank operation is performed into the received signal to undo the Synthesis Filterbank operation applied in transmission (both operations are the key components of the PHYDYAS project). Then the signal passes to the channel compensation module. In this module, the channel estimation process has to take into account the transmultiplexer response of the applied bank of filters [18]. The channel estimation is obtained by means of a two-dimensional linear interpolation based on the channel behavior at the pilot positions. Finally, the equalization is performed using the ZF criterion. The applied channels are ITU-R channel Vehicular A (VEHA) or Pedestrian B (PEDB), both presented in [21]. At the demodulation stage, the signal is postprocessed to undo the Offset QAM (OQAM) modulation inserted in the FBMC architecture and the modulated transmitted symbols are obtained. Fig. 4 depicts the performance of the receiver when demodulating QPSK symbols and a VEHA channel is applied.

## V. CONCLUSIONS

In this paper, the uPHYLA software framework to develop communication systems is presented. The use of software engineering techniques in the conception, implementation and maintenance of uPHYLA provides the framework with desirable qualities such as flexibility and modularity. This framework makes more straightforward to researchers the testing of new algorithms, accelerating the transition from simulation to practical demonstration with real signal and with the simplicity of using a common programming language such as C++. uPHYLA framework has been proven effective in the development of an UL FBMC receiver with MIMO capabilities, which includes present novelties in the field of multicarrier communications. This demonstration together with the hardware evolution in next years shows that SDR technologies are a real alternative to build flexible testbeds implementing communication systems with high demands on bandwidth to accommodate high data rate transmission.

### ACKNOWLEDGMENTS

The authors would like to thank Francisco Rubio, Antonio Pascual Iserte and Miquel Payaró for his wise recommendations, help and advice in technical issues related to understanding of multicarrier communication systems and hardware configuration. This work has been supported by the European Comission in the framework of the FP7 project PHYDYAS (contract number INFSO-ICT-211887).